\documentclass[fleqn,usenatbib]{mnras}

\usepackage{newtxtext,newtxmath}

\usepackage[T1]{fontenc}

\DeclareRobustCommand{\VAN}[3]{#2}
\let\VANthebibliography\thebibliography
\def\thebibliography{\DeclareRobustCommand{\VAN}[3]{##3}\VANthebibliography}

\usepackage{graphicx}	
\usepackage{amsmath}	

\title[Water vapour transit ambiguities]{Water Vapour Transit Ambiguities for Habitable M-Earths}

\author[Macdonald et al.]{Evelyn Macdonald$^{1}$\thanks{E-mail: evelyn.macdonald@mail.utoronto.ca}
Kristen Menou,$^{1,2,3}$
Christopher Lee,$^{1}$
and Adiv Paradise$^{3}$
\\
$^{1}$Department of Physics, University of Toronto, Toronto, ON, Canada M5S 1A7\\
$^{2}$Department of Physical and Environmental Sciences, University of Toronto, Scarborough, ON, Canada M1C 1A4\\
$^{3}$David A. Dunlap Department of Astronomy and Astrophysics, University of Toronto, ON, Canada ON M5S 3H4
}

\date{Accepted XXX. Received YYY; in original form ZZZ}

\pubyear{2024}

\begin{document}
\label{firstpage}
\pagerange{\pageref{firstpage}--\pageref{lastpage}}
\maketitle

\begin{abstract}
We have shown in a recent study, using 3D climate simulations, that dayside land cover has a substantial impact on the climate of a synchronously rotating temperate rocky planet such as Proxima Centauri b. Building on that result, we generate synthetic transit spectra from our simulations to assess the impact of these land-induced climate uncertainties on water vapour transit signals. We find that distinct climate regimes will likely be very difficult to differentiate in transit spectra, even under the more favourable conditions of smaller planets orbiting ultracool dwarfs. Further, we show that additional climate ambiguities arise when both land cover and atmosphere mass are unknown, as is likely to be the case for transiting planets. While water vapour may be detectable under favourable conditions, it may be nearly impossible to infer a rocky planet's surface conditions or climate state from its transit spectrum due to the interdependent effects of land cover and atmosphere mass on surface temperature, humidity, and terminator cloud cover.

\end{abstract}




\section{Introduction}

Temperate rocky planets orbiting M-dwarfs, or M-Earths, may have climates suitable for hosting life. JWST has started to produce transit spectra of small planets \citep{Greene2023,Zieba2023}, with more expected in the coming years. These spectra will provide some clues about the atmospheres -- or lack thereof -- of these planets, but their surfaces will be difficult to observe due to instrumental limitations. However, surface conditions are known to affect a planet's climate and habitability (e.g. \citealt{Shields2018, Lewis2018, Rushby2020, Salazar2020}). In particular, we showed in \citet{Macdonald2022} that the configuration and amount of land on a synchronously rotating M-Earth can significantly affect its humidity and temperature. We will now attempt to map these land-related climate differences to differences in the corresponding transit spectra.

In this paper, we use a general circulation model (GCM) combined with a radiative transfer model to generate synthetic transit spectra over a large parameter space for synchronously rotating, habitable-zone M-Earths. We vary the planet's land fraction, land configuration, and atmosphere mass, all of which will be difficult to independently measure. We find that while the planet's temperature and humidity are heavily dependent on the parameters varied, there are significant degeneracies between climate states in synthetic transit spectra, especially in the presence of clouds. 

\section{Methods}

\subsection{General circulation model}

We use ExoPlaSim \citep{Paradise2022}, a fast, intermediate complexity GCM which is able to simulate the climates of a diverse range of habitable planets. Our simulations are separated into three groups, summarized in Table \ref{tab:groups} and described below. All are synchronously rotating.

Group A is our ExoPlaSim simulations from \citet{Macdonald2022}, which have the parameters of Proxima Centauri b \citep{Anglada-Escude2016}. These simulations fall under two landmap classes: substellar continent (SubCont), with a circular continent at the substellar point and ocean covering the nightside and the rest of the dayside, and substellar ocean (SubOcean), with land everywhere except for a circular dayside ocean centred at the substellar point (Fig. 1 of \citealt{Macdonald2022}). The land fraction is varied from 0 to 100\% dayside land cover for both landmap classes. 

Group B are variations on Group A. Parameters are varied one at a time and land cover is systematically varied in each configuration, as in Group A. Group A and B climates are summarized in Figs. 7 and 10 of \citet{Macdonald2022}. emperature and water vapour are highest when ice-free ocean area is maximized, which occurs at low land fraction for SubCont climates and at partial dayside land cover for SubOcean climates, similar to the 1 bar Group C trends seen in Figure \ref{fig:climates} below.

Group C is a new set of simulations optimized for larger atmosphere transit depth. We use a smaller star to increase the planet-star size ratio, and a smaller planet to increase the scale height of its atmosphere while keeping the planet realistic, as per the mass-radius relation of \citet{Otegi2020}. We also shorten the planet's synchronous rotation period in a physically consistent way, such that it receives the same flux as Proxima Centauri b despite its cooler star. We use the SubCont and SubOcean landmap classes described above, again systematically varying the dayside land fraction.  

\begin{table}
\begin{tabular}{c||c|c|c}
\centering
Group & A & B & C \\
\hline
\hline
Radius (R$_\oplus$) & 1.12 & 1.12 & 0.646 \\
Gravity (m/s$^2$) & 10.9 & 10.9 & 4.7 \\
Period (days) & 11.186 & 11.186 & 4.96 \\
pN$_2$ (bar) & 1 & 1 & 0.5, 1, 2, 4, 6, 8 \\
pCO$_2$ (millibar) & 1 & 1, 100 & 1 \\
Stellar temperature (K) & 3000 & 2600, 3000, 3500 & 2600 \\
Stellar radius (R$_\odot$) & 0.13 & 0.13 & 0.1 \\
Stellar flux (W/m$^2$) & 881.7 & 700, 881.7 & 881.7 \\
Resolution (lat$\times$lon) & $96\times192$ & $96\times192$ & $64\times128$  
\end{tabular}
\caption{Description of simulations.}
\label{tab:groups}
\end{table}

\subsection{Radiative transfer model}

\begin{figure}
    \centering
    \includegraphics[width=\columnwidth]{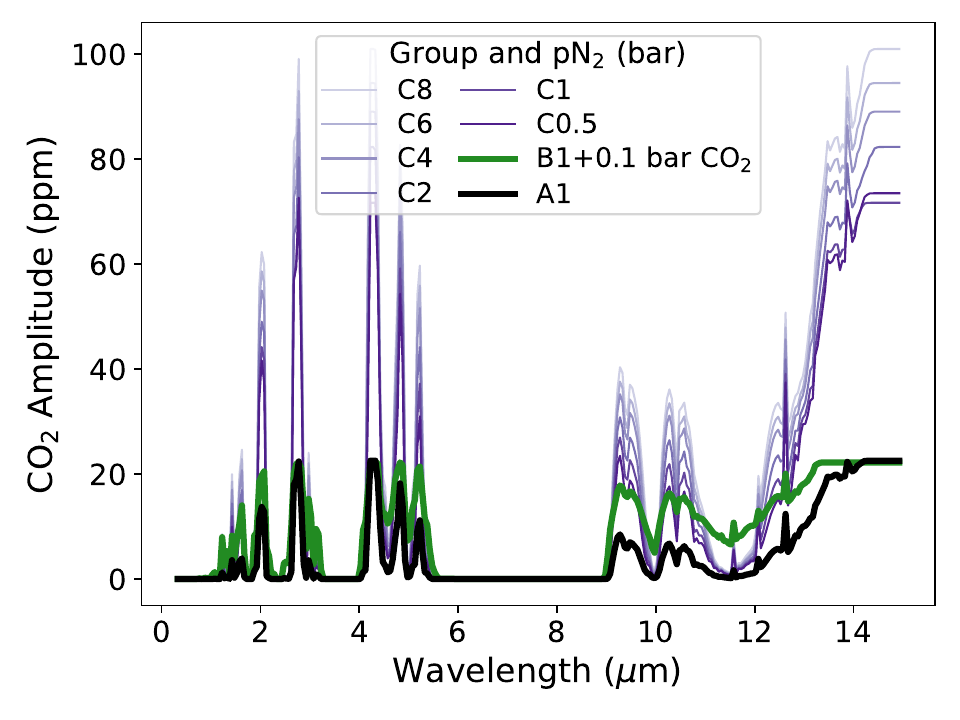}
    \caption{CO$_2$ transit spectra for aquaplanets from Group C with varying pN$_2$ (purple), Group A (black), and Group B with 0.1~bar CO$_2$. Although the Group B simulation has two orders of magnitude more CO$_2$, its transit amplitude is smaller than those of the Group C simulations due to the geometries of the systems. However, note that our GCM simulations only include a troposphere; if CO$_2$ is well-mixed on a real planet, spectral features could extend much higher.}
    \label{fig:co2spectra}
\end{figure}

To generate synthetic transmission spectra from our GCM simulations, we use the radiative transfer model petitRADTRANS \citep{Molliere2019}, which uses opacity data from the Exomol database \citep{Tennyson2016}. For each ExoPlaSim simulation, we use pressure, temperature, and specific humidity profiles to calculate transmission through each column of the atmosphere along the terminator. We disregard other gases in order to emphasize water vapour differences between models; water vapour can safely be isolated from the spectrum in this way because the effects of different molecules are additive to first order. We construct the planet's water vapour transit spectrum as the average of all of the terminator columns, weighted by cross-sectional area (Figure \ref{fig:profiles}). For comparison, we also show dry CO$_2$ spectra in Figure \ref{fig:co2spectra}.

\section{Results}

\subsection{Groups A and B}

\begin{figure*}
    \centering
    \includegraphics[width=\textwidth]{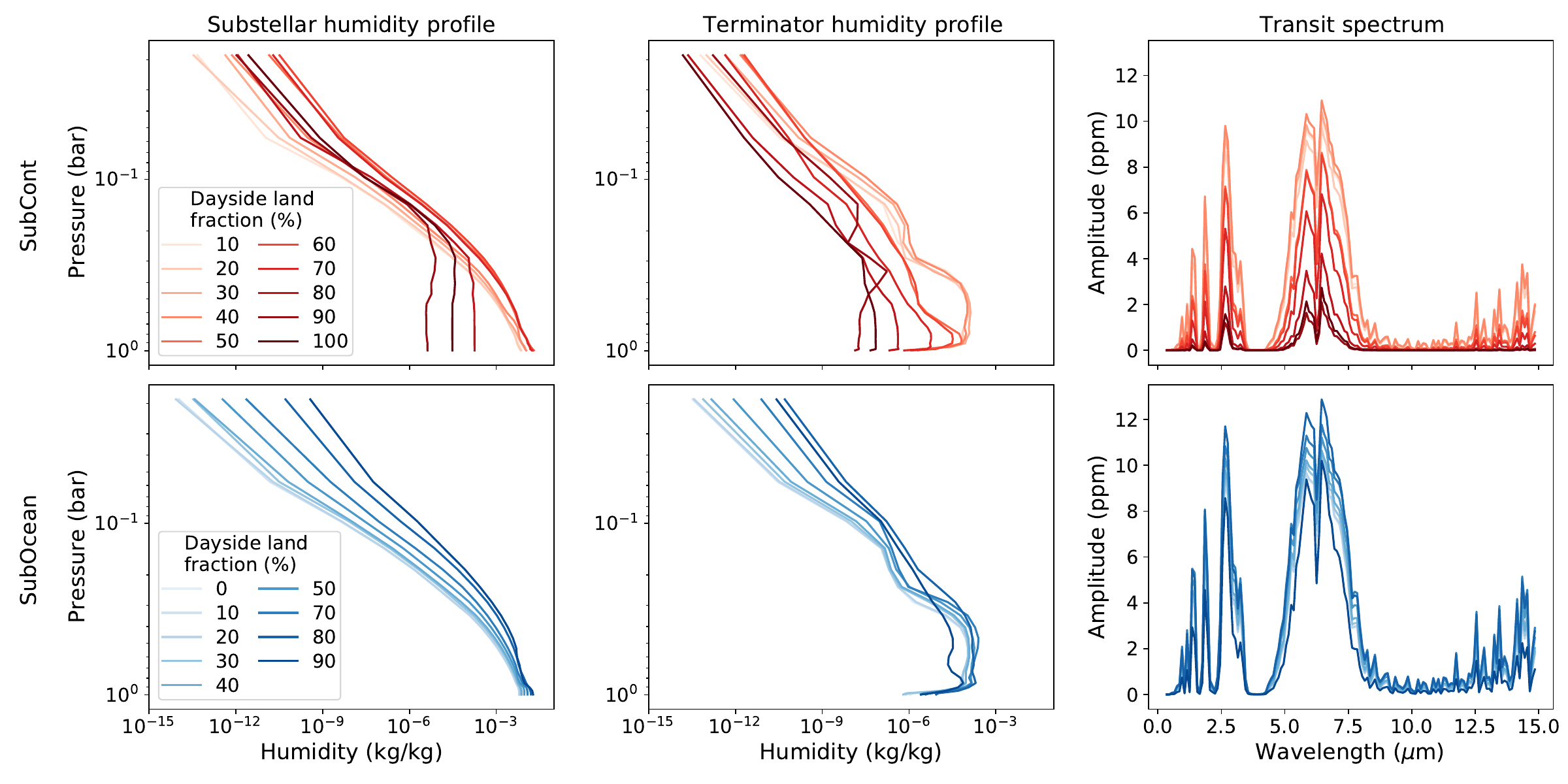}
    \caption{Left to right: substellar and terminator specific humidity profiles, and water vapour transit spectra, for Group A SubCont (top row) and SubOcean (bottom row) simulations. Low-land-fraction SubCont models, which also have sizeable ice-free oceans, resemble SubOcean models in transit. Higher-land-fraction SubCont models have little to no ice-free ocean, and their water vapour transit depth decreases steadily with increasing land fraction. There is no clear separation between SubCont and SubOcean spectra, or between spectra of SubCont models with and without ice-free ocean.}
    \label{fig:profiles}
\end{figure*}

Substellar and terminator-averaged specific humidity profiles and cloud-free synthetic water vapour transit spectra for Group A are shown in Figure \ref{fig:profiles}. All simulations have less water at the terminator than at the substellar point. There is significantly more variation in both profiles and spectra of SubCont than SubOcean models due to the higher variability in ice-free ocean area in the former. Because the water in SubOcean simulations is centred at the warmest part of the planet, at least some of this ocean is always ice-free, so evaporation can take place. As a result, water vapour can enter the atmosphere even when the substellar ocean is small, so profiles and spectra of SubOcean simulations do not depend heavily on dayside land fraction. Low-land-fraction SubCont models have spectra that fall within the range of the SubOcean models, meaning that it would be difficult to differentiate between an ocean planet with a small substellar continent and a planet with ocean covering only the central 10\% of its dayside. For SubCont models, the substellar and terminator humidities and the amplitudes decrease steadily as the land fraction increases. The lack of clear bimodality between spectra of models with and without ice-free ocean will make it difficult to determine from a transit spectrum whether a planet has surface liquid water.

\begin{figure}
    \includegraphics[width=\columnwidth]{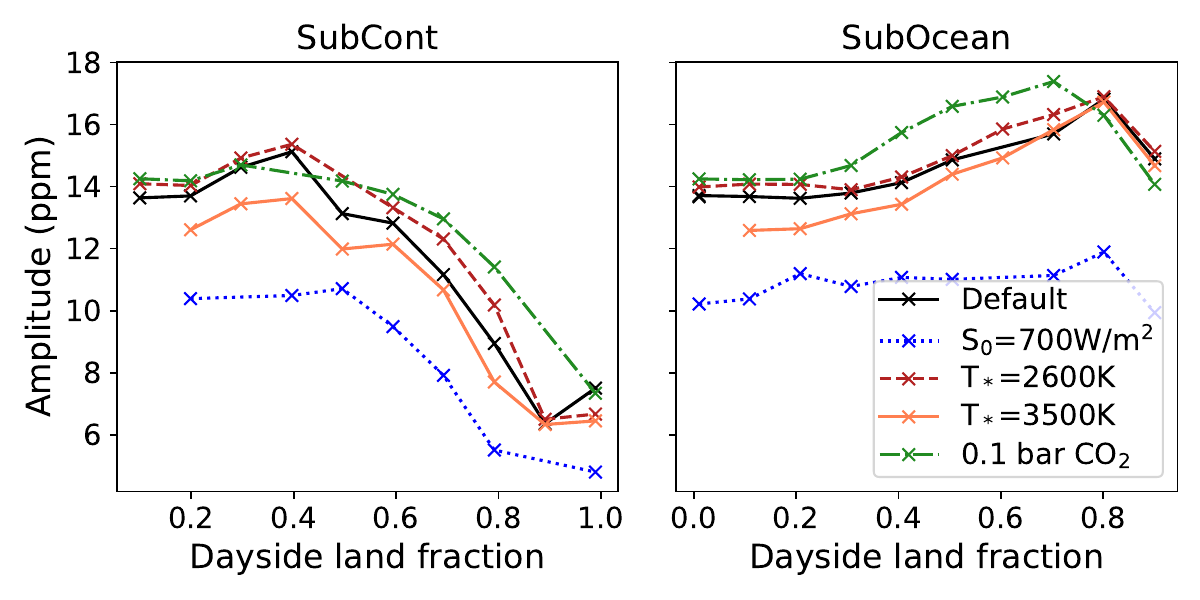}
    \caption{Spectral feature amplitude, defined as the maximum differential transit depth of the water vapour peak ($\Delta$ppm), as a function of dayside land fraction, for Groups A and B. These trends are qualitatively similar to the temperature and water vapour trends shown in \citet{Macdonald2022}; however, the diminutive scale of the differences between models will likely make it very difficult to tell these climates apart observationally with current capabilities. In particular, the noise floor of JWST is estimated at around 10~ppm \citep{Schlawin2021, Rustamkulov2022}.}
    \label{fig:fig2}
\end{figure}

To facilitate comparison between simulations, we show in Figure \ref{fig:fig2} the amplitude of the $\sim6\,\mu$m water vapour spectral feature as a function of dayside land fraction for Groups A and B. This amplitude is defined as the maximum differential transit depth ($\Delta$ppm) of the spectrum. We find that amplitude has a similar land fraction dependence to temperature and water vapour (\citealt{Macdonald2022}, Figs. 7 and 10) in these simulations. This implies that the substantial climate differences caused by landmap changes are recovered in transit spectra, albeit at too small of a scale for detection with existing instruments.

\subsection{Group C}

\begin{figure}
    \includegraphics[width=\columnwidth]{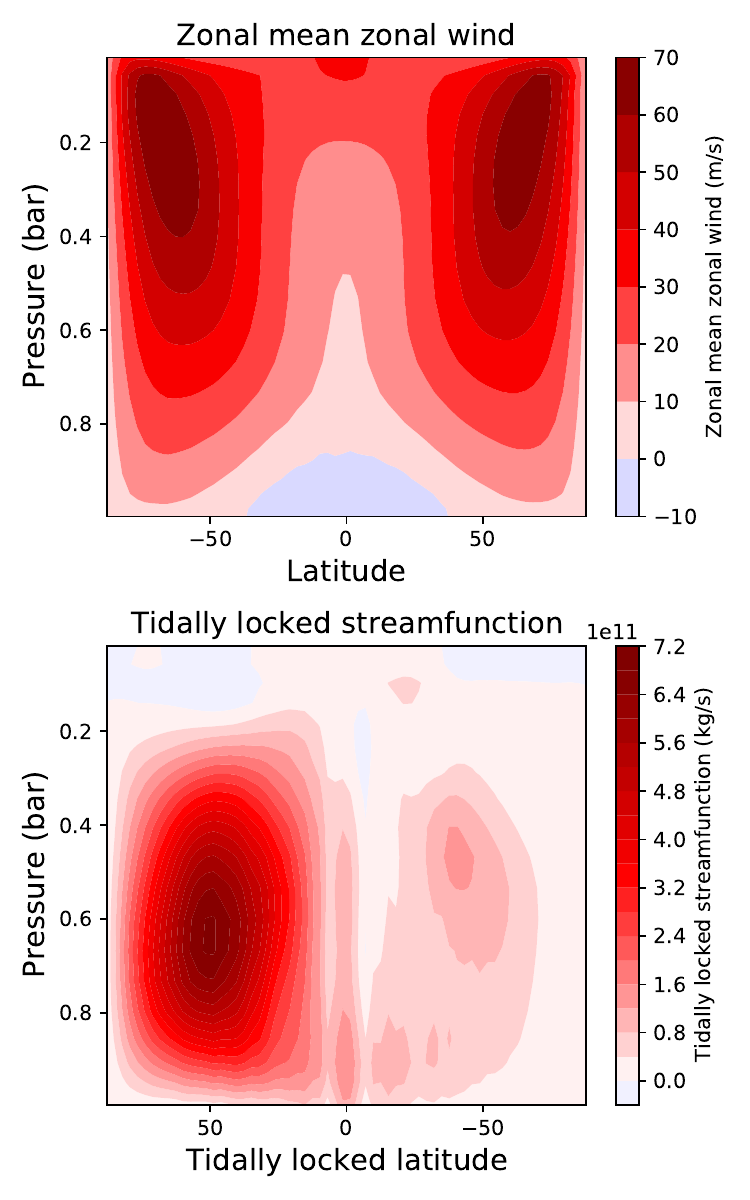}
    \caption{Circulation of a 0.2~M$_\oplus$ aquaplanet with a 1~bar N$_2$ atmosphere orbiting a 2600~K M-dwarf. Top: zonal mean zonal wind. Bottom: tidally locked streamfunction, in tidally locked coordinates where 90$^\circ$, 0$^\circ$, and -90$^\circ$ represent the substellar point, the terminator, and the antistellar point, respectively \citep{Koll2015,Hammond2021,Paradise2022}. This circulation regime, featuring two mid-latitude zonal jets and a general overturning circulation from dayside to nightside, is representative of most Group C simulations.}
    \label{fig:circulation}
\end{figure}

To explore a more observationally favourable system for water vapour detection, we include a new set of simulations of a 0.2~M$_\oplus$ planet orbiting a 2600~K M-dwarf (Group C in Table \ref{tab:groups}). Note that 0.2~M$_\oplus$ is approximately 2 Mars masses, and larger than multiple Kepler planets, so this is a reasonable size for a terrestrial planet with an atmosphere. The planet's small size is chosen to maximize the amplitude of spectral features, since the atmosphere's scale height is inversely related to the planet's surface gravity. In most cases, the shorter rotation period does not cause a qualitative shift in dynamic regime relative to Groups A and B. Figure \ref{fig:circulation} shows the zonal mean zonal wind and tidally locked streamfunction for a sample simulation. This streamfunction describes the circulation in tidally locked coordinates, with 90$^\circ$ at the substellar point, 0$^\circ$ at the terminator, and -90$^\circ$ at the antistellar point \citep{Koll2015,Hammond2021,Paradise2022}. Our simulated planets are Rhines rotators \citep{Haqq-Misra2018} with two zonal jets and overturning circulation from the substellar point toward the nightside. Most precipitation in this circulation regime falls near the substellar point regardless of land configuration.

\begin{figure*}
    \centering
    \includegraphics[width=\textwidth]{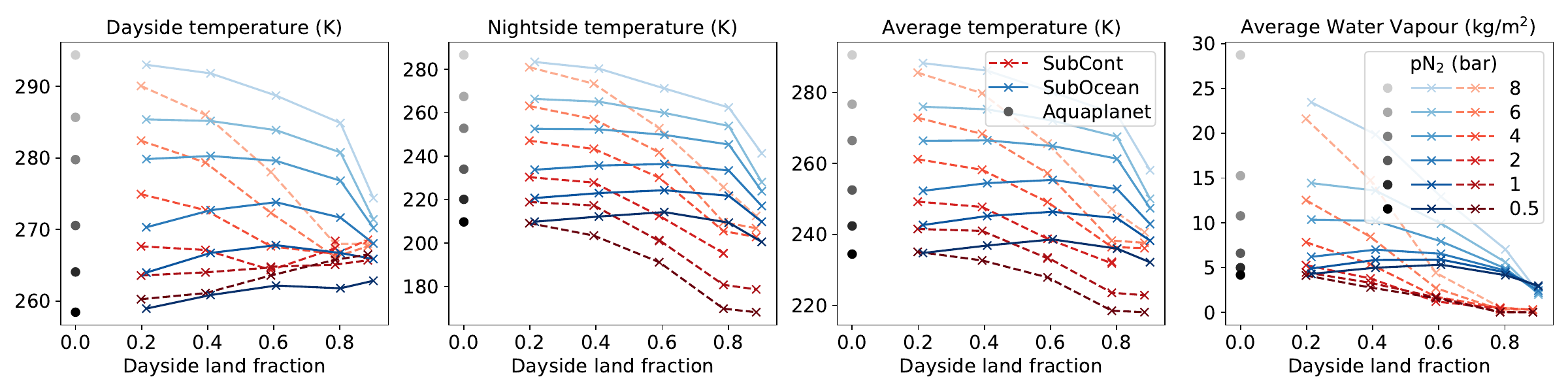}
    \caption{Left to right: dayside, nightside, and globally averaged surface temperature, and globally averaged water vapour, for Group C. The climate trends are similar to Groups A and B, with SubOcean simulations exhibiting higher temperature and humidity than SubCont simulations at partial dayside land cover for a given pN$_2$. Many of these curves intersect, meaning that different combinations of land cover and pN$_2$ can result in similarities between climates.}
    \label{fig:climates}
\end{figure*}

Group C climate trends are shown in Figure \ref{fig:climates}. The land fraction and configuration dependence is qualitatively similar to that of Groups A and B, with a shift to higher temperatures and humidities as pN$_2$ increases. This pN$_2$-induced warming is expected for cool stars, because warming occurs due to pressure broadening, but the cooling effect of Rayleigh scattering is minimal since the Rayleigh scattering cross section scales as $\lambda^{-4}$, and thus is very small at the wavelengths at which M-dwarfs emit most of their light \citep{Paradise2021,Paradise2022}.

\begin{figure*}
    \includegraphics[width=\textwidth]{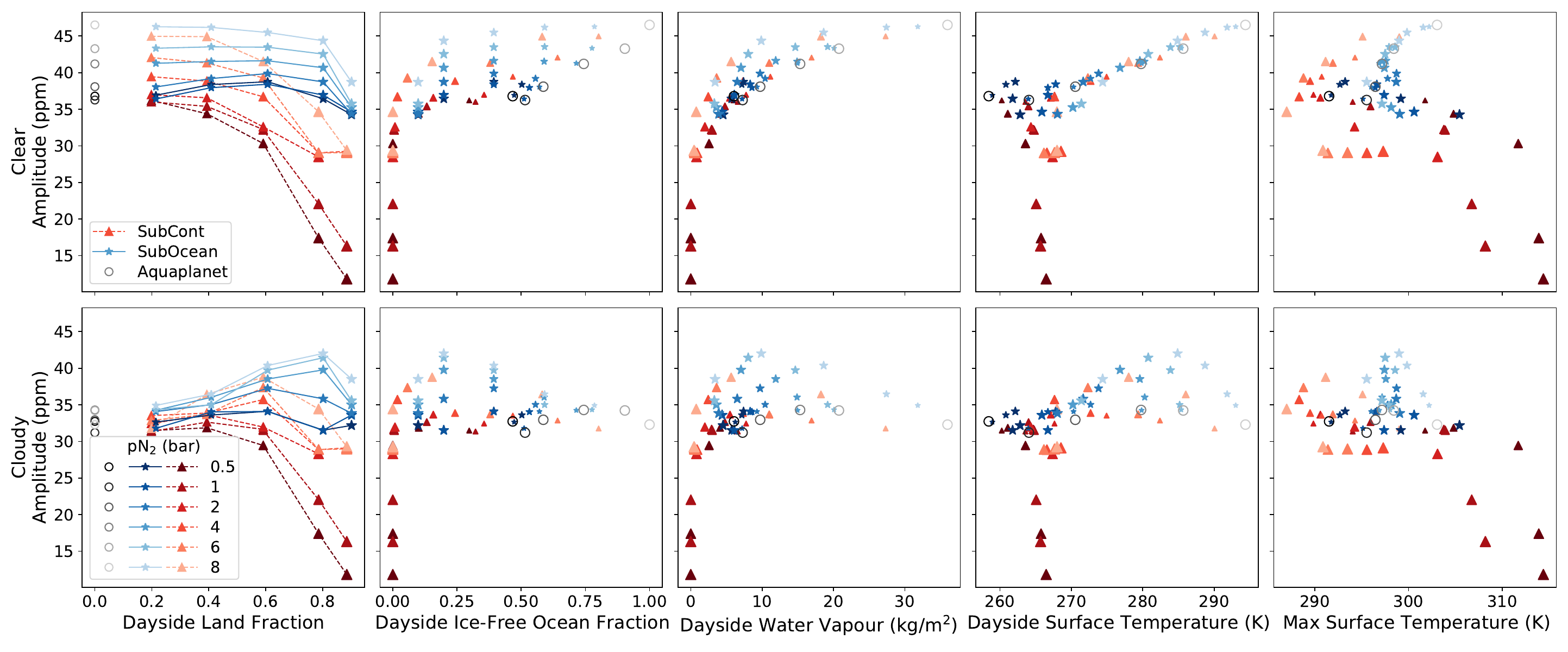}
    \caption{Maximum transit amplitude (y-axis) of the water vapour peak for Group C simulations in clear sky (top row) and cloudy (bottom row). Left to right (x-axis): dayside land fraction, ice-free ocean fraction, atmospheric water vapour, average surface temperature, and maximum surface temperature. Marker sizes are proportional to dayside land fraction. High-pN$_2$, low-land-fraction SubOcean models have the largest water vapour amplitudes when clouds are ignored. Clouds make water vapour more challenging to detect overall. Their effect is particularly pronounced for these moist atmospheres, such that water vapour transit amplitude only weakly correlates to climate variables.}
    \label{fig:amp900}
\end{figure*}

The top left panel of Figure \ref{fig:amp900} shows the amplitude of the water vapour transit signal as a function of dayside land fraction for Group C. The trends are qualitatively similar to the Group A and B trends (Figure \ref{fig:fig2}), but with significantly larger amplitudes. Each of land fraction, land configuration, and pN$_2$ can vary the expected peak amplitude by more than 10~ppm, and their combined effect brings this variation closer to 20~ppm, or more in the case of very dry, high-land-fraction, low-pN$_2$ planets.

\subsection{Clouds}

We include clouds in our petitRADTRANS spectra by prescribing profiles using the cloud liquid water field of our ExoPlaSim outputs and a temperature-dependent cloud particle size parameterization from \citet{Edwards2007}. The bottom left panel of Figure \ref{fig:amp900} shows the effect of clouds on the maximum water vapour transit amplitude for Group C simulations. Clouds obscure spectral features, especially for simulations with low dayside land fraction and high pN$_2$, which have the most water vapour and the cloudiest terminators. As a result, some high-land-fraction spectra display water vapour the most prominently, despite not actually having the moistest atmospheres.

The four right columns of Figure \ref{fig:amp900} show water vapour amplitude as a function of dayside ice-free ocean fraction (defined as the fraction of the dayside that is neither land nor ice), average water vapour and surface temperature, and maximum surface temperature. Without clouds (top row), the amplitude is positively correlated to the first three, except in the driest cases; maximum surface temperature is more obscure because dry models have larger day-night temperature contrasts. The bottom row shows the same variables when clouds are included. Clouds largely obscure the correlations, but do not make water vapour entirely undetectable.

\section{Discussion and Conclusions}

Using ExoPlaSim and petitRADTRANS, we have generated synthetic transit spectra for a range of synchronously rotating, potentially habitable planets with a range of land configurations. We have found that for the Proxima b-sized planet from \citet{Macdonald2022}, terminator water vapour is consistently high in simulations with substellar oceans, and is much more dependent on land fraction in models with substellar continents. The water vapour amplitudes of the corresponding spectra correlate to globally averaged surface temperature and atmosphere water vapour content; however, there is no clear separation between the transit spectra of the two landmap classes or of planets with and without ice-free ocean, and the transit signal is weak in all cases. 

We have shown that these relationships hold for a more observationally favourable system with a smaller planet orbiting a smaller star. The addition of atmosphere mass as an independent variable creates further ambiguity. There is a trend toward higher temperature, humidity, and transit depth with increasing pN$_2$ which is largely obscured by the inclusion of clouds. The differential transit depth of our simulations ranges from 12-47~ppm in the clear case and 12-42~ppm when clouds are included, with all but the driest planets above 28~ppm. Aquaplanets and low-land-fraction planets are spread widely in the upper part of the range in the former case and clustered around 30-35~ppm in the latter because of their cloudier terminators. 

The interpretation of exoplanet transit spectra will depend heavily on clouds. Terminator cloud cover, which significantly affects a planet's spectrum, is difficult to measure because a cloud deck can be indistinguishable from the planet's surface; consequently, the altitude of the minimum transit depth is uncertain. Further, clouds are a major source of modelling uncertainty. \citet{Komacek2020} found using ExoCAM simulations that clouds can cause a drastic reduction in spectral feature amplitude. \citet{Wolf2022ExoCAM} note that ExoCAM clouds are sensitive to changes in the model's parameterizations. \citet{Sergeev2022, Fauchez2022} found significant differences in cloud-related climate variables between four GCMs (ExoCAM, ROCKE-3D, LMD-G, and UM) in simulations of TRAPPIST-1e. ROCKE-3D generally produced the most cloud cover, but ExoCAM had the most cloud liquid water on the nightside and the highest-altitude terminator clouds, which resulted in the largest impact on transit spectra. These differences in cloud and convection parameterizations resulted in inter-GCM cloud-related uncertainties of up to 50\% in the number of transits required to detect molecules in the transmission spectrum of TRAPPIST-1e. By running ExoPlaSim simulations of these cases and the slower-rotating planet of \citet{Yang2019}'s intercomparison, \citet{Paradise2022} found that ExoPlaSim was broadly consistent with these other GCMs, but produced lower nightside cloud cover than ExoCAM. It is therefore possible that ExoPlaSim is underestimating terminator cloudiness, in which case our results would represent an optimistic estimate of water vapour detectability. 

Our synthetic transit spectra do not show the uncertainties that will be present in JWST data, which will have contributions from photon noise (e.g., \citealt{Cowan2015}) and stellar variability. The noise floor of JWST, estimated at around 10~ppm \citep{Schlawin2021, Rustamkulov2022}, will add to the challenge of detecting small spectral features.

The parameter space explored in this study covers a large range of potentially habitable climates. Although water vapour and other spectral features may be detectable in some cases, the combination of unknown land fraction, land configuration, and atmosphere mass will make it difficult or impossible to precisely infer an M-Earth's climate or surface conditions from its transit spectrum in the near future. Imperfectly modelled clouds add considerable uncertainty. More work is needed to improve cloud models so that ambiguities in future data can be better understood.

\section*{Acknowledgements}

EM is supported by a Natural Science \& Engineering Research Council (NSERC) Post-Graduate Scholarship and by the University of Toronto Department of Physics. KM is supported by NSERC. CL is supported by the Department of Physics. We would like to thank an anonymous reviewer whose comments improved this paper.

The University of Toronto, where most of this work was performed, is situated on the traditional land of the Huron-Wendat, the Seneca, and the Mississaugas of the Credit. This study made substantial use of supercomputing resources; although most of the energy used comes from low-carbon sources, we acknowledge that building and operating these facilities has a significant environmental impact.

\section*{Data Availability}

The simulation outputs for this study and the files needed to reproduce them are available in Borealis repositories \citep{Macdonald2021,Macdonald2024data}, as per \citet{Paradise2020}. The ExoPlaSim source code is available at \url{https://github.com/alphaparrot/ExoPlaSim/}. The petitRADTRANS source code is available at \url{ https://gitlab.com/mauricemolli/petitRADTRANS}.



\bibliographystyle{mnras}
\bibliography{references} 



\bsp	
\label{lastpage}
\end{document}